# Generation of high peak power, segmented and Bessel beams of tunable range without on-axis intensity modulation

A. Srinivasa Rao,* G. K. Samanta

Photonic Sciences Lab, Physical Research Laboratory, Navarangpura, Ahmedabad 380009, Gujarat, India
*Corresponding author: asvrao@prl.res.in

**We propose and experimentally demonstrate a novel experimental scheme to generate high peak power, segmented, smooth, zero-order Bessel beams with tunable range. Illuminating the axicon with hollow Gaussian beams (HGBs) of different orders we have generated Bessel beams of varying range at different positions away from the axicon. The presence of dark core at the center of the HGBs removes the effect of imperfection in the axicon tip. As a result, the entire power of the input beam is transformed into zero-order Bessel beam without any on-axis intensity modulation. We observe the decrease in range and increase in peak power of the zero-order Bessel beam with the order of HGBs. Controlling the superposition of the HGBs of different orders to the axicon we have demonstrated the increase in the range of the Bessel beam. The current technique can also produce Bessel beams of different intensity distribution including single peak or multiple peak Bessel beams. Using single-pass second harmonic generation in nonlinear crystals of different lengths we have further verified the increase of peak power of the Bessel beam with the order of the HGBs and also the increase in the range of the Bessel beam due to the superposed HGBs. The generic experimental scheme can be used at different wavelengths and timescales (continuous-wave to ultrafast).**

Diffraction, a fundamental phenomenon intricately liked to all classical waves, causes the light beams to spread along propagation. In 1987, Durnin showed the possibility of generating diffraction-free optical beams over finite range along beam propagation [1]. The intensity profile of these finite ranges, diffraction-free beams can be described by Bessel function and known as Bessel beams [2-4]. The Bessel beam is characterised with its order. The zero-order Bessel beam has intensity distribution with bright central lobe surrounded by infinite rings separated at π phase difference in the radial direction [2], and the higher order Bessel beams have the doughnut shaped central lobe and carry orbital angular momentum (OAM) [3] per photon. Due to the presence of central bright lobe, the zero-order Bessel beams have drawn a great deal of interest for variety of applications in science and technology including accelerating charged particles [5], atomic guiding [6], optical coherence tomography [7], precision alignment and power transport [8], and material processing [9].

A variety of techniques based on annular slit [1], computer-generated holograms [4], meta-surfaces [10], digital micro-mirror devices [11], and axicon [3], have been used to generate Bessel beams. Due to high damage threshold and broad wavelength coverage, the axicons are of preferred choice to generate high power/energy zero-order Bessel beams particularly important for applications in nonlinear optics or laser processing. However, the roundness in the tip of the axicons arising from the manufacturing defects results zero-order Bessel beam with strong on-axis oscillations in the spatial intensity profile, which is undesired for practical applications [12]. Efforts have been made to remove such on-axis oscillation with the help of Fourier filter [13] or by blocking the beam passing through the tip of the axicon [14], but at the cost of lower Bessel beam power. Additionally, the axicon produces zero-order Bessel beam right after the axicon. As a result, there is no control in generating the Bessel beam at a desired axial position along propagation and also in the tunability of the Bessel beam range. Very recently, the combination of axicon and variable annular slit has produced segmented zero-order Bessel beam with controlled range at different distances along the propagation [14]. However, the use of annular aperture reduces the overall power of the zero-order Bessel beam and also the diffraction effect restricts the effective range of the Bessel beam.

Since the annular aperture has close resemblance with the optical beams of doughnut intensity distribution, a dark center enclosed by a bright ring in the beam cross section, using such beams one can, in principle, generate Bessel beam without on-axis intensity modulation. As such, optical vortices, having doughnut intensity distribution with zero intensity at the beam axis, have been used to produces higher order Bessel beam [3]. However, the generation of zero-order Bessel beam requires doughnut shaped beams with plane wavefront. Here, we propose and experimentally demonstrate a novel scheme based on axicon producing high power segmented smooth zero-order Bessel beams with tunable range. Illuminating an axicon with hollow Gaussian beams (HGBs), a special class of doughnut shaped beams carrying plane wavefront, of different orders we showed the control in the appearance of the zero-order Bessel beam along the propagation distance. We also predicted the increase in the peak intensity of the Bessel beam with the order of the HGB and the control in the range of the Bessel which further verified by second harmonic generation (SHG) process.

The transverse electric field distribution of a HGB of power, $P$, at $z=0$ plane can be represented as [15, 16],

$$E_l(r, \theta) = \left(\sqrt{\frac{2^l P}{w_a^2 l!}}\right)\left(\frac{r^2}{w_a^2}\right)^l \exp\left(\frac{-r^2}{w_a^2}\right) \quad (1)$$

where, $l$ is the order of the HGB and $w_a$ is the waist radius of the Gaussian beam hosting the HGB. Solving the Fresnel diffraction integral using stationary phase method [3, 4] for the axicon illuminated by the HGB having transverse electric field distribution given by Eq. (1), we can derive the intensity distribution of the output beam in the form of

$$I(r, z) \approx I(z) J_0^2(k_r r) \quad (2)$$

where,

$$I(z) = \frac{2^l P}{l! \pi w_a^2}\left(\frac{z}{z_{max}}\right)^{4l+1} \exp\left(\frac{-2z^2}{z_{max}^2}\right) \quad (3)$$

here, $z_{max}=w_a/(n-1)\alpha$ is the range of the beam, $n$ and $\alpha$ are the refractive index and base angle of the axicon, respectively. The Eq. (2) resembles the intensity distribution of a zero-order Bessel beam confirming the generation of zero-order Bessel beam by illuminating axicon with HGBs. The dependence of longitudinal intensity distribution, $I(z)$, of the Bessel beam on the orders, $l$, of the HGBs, as shown by Eq. (3), enables the control in the overall intensity distribution of the Bessel beam. For $l=0$, $I(z)$ represents the axial intensity of zero-order Bessel beam generated by the Gaussian beam. Here, $z=0$, represents the axicon position. The experimental scheme is pictorially depicted in Fig. 1(a). As evident, for a fixed axicon, the start, $z_{start}$, and end, $z_{end}$, of the Bessel beam along beam propagation are determined by the inner radius, $r_a$, and outer radius, $r_b$, of the annular intensity ring of the HGB. The $z_{start}$, and $z_{end}$ can be represented as, $z_{start}=r_a/(n-1)\alpha$ and $z_{end}=r_b/(n-1)\alpha$ and the range the Bessel beam is given as, $\Delta z = z_{start} - z_{end}$. As the annular ring radius of the HGB depends on its order, $l$, we can actively control the positions $z_{start}$, and $z_{end}$ and the range of Bessel beam by illuminating the axicon with HGBs of different orders.

The schematic of the experimental setup is shown in Fig. 1. A 50 W, cw Yb-fiber laser [17] (IPG Photonics, YLR-50-1064), delivering single-frequency, linear polarized radiation at 1064 nm in $TEM_{00}$ spatial profile ($M^2 < 1.07$), is used as the fundamental source. The laser power to the experiment is controlled by using a combination of half wave plate ($\lambda/2$) and a polarizing beam splitter (PBS1) cube. Using the $\lambda/2$ plate and PBS2 we have divided the Gaussian pump beam into two arms (arm 1 and arm 2). Using the spiral phase plates (SPPs), SPP1 and SPP2 of phase-winding corresponding to vortex orders, $l_{OV}=1$ and $l_{OV}=2$, respectively, PBS (PBS3 and PBS4), $\lambda/2$ and mirrors, M, we can convert the Gaussian beam of each arm (arm 1 and arm 2) into HGBs of orders $l= 1, 2$ and 3. The working principle of HGB generation [18] can be understood as follows. The horizontal (H) polarized Gaussian beam transmitted through the PBS2 acquires vortex order, $l_{OV}=+1$ (say) while passing through the SPP1. The H polarized vortex beam transmits through the PBS3 and transformed into vertical (V) polarized vortex beam on transmission through the $\lambda/2$ having an optic axis at 45° with respect vertical direction. On reflection by the mirrors, M, the vortex beam changes it's sign from $+l_{OV}$ to $-l_{OV}$ and vice versa. Since the vortex beam undergoes four reflections (three mirror reflections and one reflection by PBS3), the return vortex beam has the same order and sign, $l_{OV}=+1$. Since the phase variation of the SPP1 as observed by the return vortex is opposite, the return beam loses its azimuthal phase and transformed into HBG of order, $l=1$. The same principle is maintained for the HGB in arm 2. The HGBs from both the arms are recombined using the PBS2. Two mirrors, M, of arm 2 are placed on a delay stage to adjust the phase and beam divergence between the HGBs generated from both the arms. The axicon of apex angle 176° transforms the HGBs into Bessel beam. A 30 mm long, 1 x 2 mm² in aperture MgO doped periodically poled lithium tantalate (MgO:CLT) with a grating of period ~7.97 µm and a 50 mm long, 1 x 2 mm² in aperture MgO doped periodically poled lithium niobate (MgO:CLN) of grating period 6.94 µm are used for frequency doubling characteristics of the Bessel beam. The wavelength separator, S, extracts green beam from the fundament.

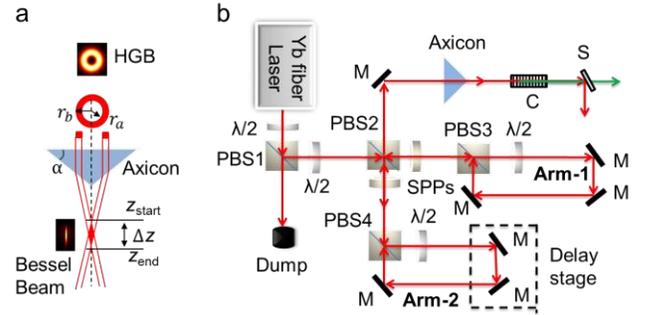

Fig. 1. (a) Schematic representation of zero-order Bessel beam generation using HGBs. (b) Experimental setup for the generation of zero-order Bessel beam. $\lambda/2$, half-wave plate; PBS1-4, polarizing beam splitter cube; SPP1-2, spiral phase plates; M, mirrors; Dump, power dumper; C, nonlinear crystal; S, wavelength separator.

We have studied the characteristics of the Bessel beams generated by the HGBs as input to the axicon and compared with the characteristics of the Bessel beam generated from Gaussian beam. The results are shown in Fig. 2. While HGBs can be generated from both arm 1 and arm 2, we have generated HGBs of orders $l= 1, 2$ and 3 using the SPP1 and SPP2 in arm 1 and blocked the beam from arm 2 for the present study. The first column, (a-d), of Fig. 2, shows the transverse intensity distribution of the HGBs measured before the axicon using a CCD camera with pixel size ~4.4 x 4.4 µm². As expected, the Gaussian beam has high intensity at the centre of the beam and the HGBs of all orders have doughnut shaped intensity distribution with zero intensity at the beam centre. It is also evident that the diameter of the dark core of HGBs increases with its order, $l= 1, 2$ and 3. Using the experimental parameters we have calculated the intensity distribution of the Gaussian beam and the HGBs, as shown in second column, (e-h), of Fig. 2, in close agreement with the experimental results. To observe the propagation dynamics of the Bessel beam we have recorded the transverse intensity distribution of the beam at different distances, z, away from the axicon by translating the CCD camera at an interval of 2 mm along the propagation direction. The line profile along $x$-axis (or $y$-axis) of the transverse intensity distribution of the beam along z-axis shows the propagation dynamics of the Bessel beam. To avoid unwanted errors in the experimental data, we have mounted the CCD camera on a 100 cm long translational stage along the $z$-direction. Due to the mechanical constrain of the experimental setup, we cannot measure the Bessel beam profile closer than $z=20$ mm from the axicon.

The experimentally measured intensity distribution of the Bessel beams in xz-plane are shown in third column, (i-l) of Fig. 2. As evident from the third column, (i-l), of Fig. 2, the Bessel beam generated from the Gaussian beam starts right after the axicon (here, z=20 mm), however, for input HGBs, the Bessel beams appear at a distance away from the axicon. It is also interesting to note that with the increase in the order of the HGBs, the Bessel beam appears at longer distance from the axicon. Using the experimental parameters in Eq. 2 we have calculated the propagation dynamics of the Bessel beam generated from the Gaussian beam and the HGBs of different orders with the results shown in fourth column, (m-p) of Fig. 2. The theoretical results are in close agreement with the experimental results. The shift in the Bessel beam generation along the propagation distance with the increase in the order, $l$, of the HGBs can be attributed to the increase in the diameter of the doughnut intensity pattern of the HGB with its order. In addition to the shift in the Bessel beam generation, the range, $\Delta z = z_{start} - z_{end}$, of the Bessel beam decreases with the increase in the order of the input HGBs. Therefore, changing the order of the input HGBs we can control the position as well as the segment of the Bessel beam. Unlike the generation of segmented beam using annular aperture [14], here, the transformation of the entire power of the input HGBs results high power, segmented Bessel beam. On the other hand, since the tip of the axicon sees the dark region of the HGBs, any imperfection in the axicon tip does not contribute to the unwanted on-axis intensity modulation [12] of the Bessel beam.

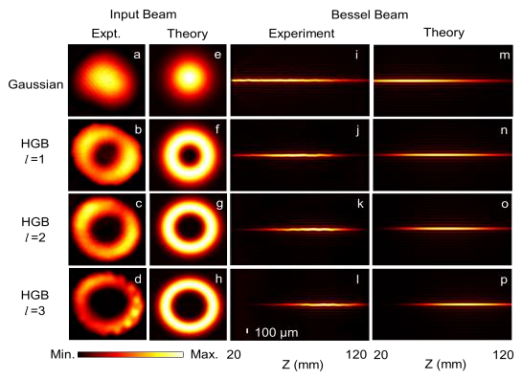

Fig. 2. (a-d) Experimental and (e-h) theoretical spatial intensity distribution of the Gaussian beam and the HGBs of different orders before the axicon. (i-l) Experimental and (m-p) theoretical intensity distribution of the Bessel beams along the propagation direction.

We have also studied the effect of imperfection in the axicon tip to the spatial intensity distribution of the zero-order Bessel beam generated by HGBs. Varying the order of the HGBs at same power before the axicon we have measured the central lobe intensity of the generated Bessel beam along the propagation distance with the results are shown in Fig. 3. For comparison, we have also measured the intensity of the central lobe of the Bessel beam generated by the Gaussian beam. As evident from the Fig. 3, the Bessel beam (blue dots) generated by the Gaussian beam carries undesired oscillation in its spatial intensity distribution confirming the imperfection in the axicon tip. However, the zero-order Bessel beams generated from the HGB of orders, $l$= 1 (red dots), 2 (green dots) and 3 (black dots) carry a smooth intensity distribution without any sign of intensity oscillation due to the imperfection in the axicon tip. Such effect can be attributed to the fact that the presence of dark core of the HGBs falling on the axicon tip nullifies the effect of imperfection of the axicon tip to the intensity distribution of the generated Bessel beams. However, we observe the decrease in the range of the zero-order Bessel beam from 95 mm for the Gaussian beam, to 76 mm, 69 mm, and 64 mm for the HGBs of orders, $l$= 1, 2, and 3, respectively. Since the range of the Bessel beam decreases with the increase in the order of the input HGBs and the entire power of the input beam to the axicon is transformed into the Bessel beam we expect an increase in the intensity of the Bessel beam. As evident from Fig. 3, the peak intensity of the Bessel beams normalized to the peak intensity of the Bessel beam generated by Gaussian, increases from 1 for the Gaussian beam to 1.46, 1.66, and 1.79 for the HGBs of orders, $l$= 1, 2, and 3 respectively. It is also interesting to note that the peak intensity positions of the Bessel beams are recorded to be at 48 mm, 62 mm, 74 mm, and 83 mm away from the axicon. It is evident from the current study that simply using the HGBs as input to the axicon and controlling the order of the HGBs, one can directly control the position of the Bessel beam generation, range, and its peak intensity.

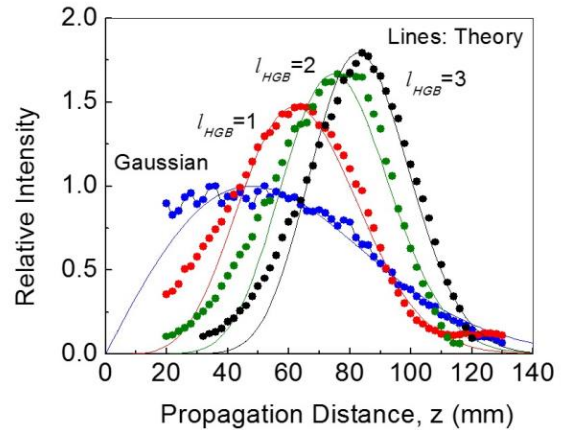

Fig. 3. Variation of intensity line profile of the central peak of Bessel beam generated from input Gaussian, and HGBs of orders, $l$= 1, 2, and 3. The intensity is normalized to the peak intensity of the Bessel beam generated from HGB of orders, $l$= 3. Solid lines are theoretical fit to the experimental results (dots).

Although the increase in peak power resulting from the decrease in the range of the Bessel beam might be beneficial for some of the applications [9], many of the applications require Bessel beams of long range. Therefore, we have also explored the possibility of extending the overall range of the Bessel beam in our experiment. Since the Bessel beams move away from the axicon with the increase in the order of the HGBs, we have combined two HGBs of different orders as input to the axicon and recorded the spatial intensity profile of the Bessel beam. The experimental results are shown in Fig. 4. Using the SPPs in arm-1 and arm-2, we have generated two independent HGBs with different orders and superposed using the PBS2 (see Fig. 1(b)). The first column, (a-c), of Fig. 4, shows the resultant intensity distribution of the beam recorded before the axicon for the beam combinations of (arm-1, arm-2) as (Gaussian, HGB, $l$= 1), (Gaussian, HGB, $l$= 2) and (HGB, $l$= 1, HGB, $l$= 2). Using the experimental parameters, we have calculated the intensity distribution of the superposed beams before the axicon, as shown in second column, (d-f), of Fig. 4, in close agreement with the experimental results. It is evident from third column, (g-i), of Fig. 4, showing the spatial intensity distribution of the Bessel beam along propagation, that the resultant Bessel beam has different ranges for the different combinations of superposed beam to the axicon. The theoretical results as presented in fourth column, (j-l), of Fig. 4, calculated using the experimental parameters show a close agreement with the experimental results. Using the intensity line profile of the central along propagation we have calculated the range of the Bessel beam to be 124 mm, 130 mm and 151 mm for the superposed beams of (Gaussian, and HGB, $l$= 1), (HGB, $l$= 1, and HGB, $l$= 2) and (Gaussian, and HGB, $l$= 2), respectively, significantly larger than the range of the Bessel beam generated from Gaussian beam (see Fig. 2). Such observation clearly proves the possibility of increasing the range of the Bessel beam by controlling the superposition of HGBs input to the axicon. Since the position of the Bessel beam can be controlled by changing the divergence and the beam size before the axicon, we have used a delay stage in arm-2 to control the superposition of the Bessel beams to form spatial intensity distribution in different shapes like single and double peaked Bessel beams. For example, one can control the separation between the peak position of the Bessel beams of HGB of arm-1 and arm-2 to produce a potential well. Although we have used only two beams, one can in principle, use beam superposition technique to couple large number of beams and generate potential well in different shapes and larger range Bessel beams.

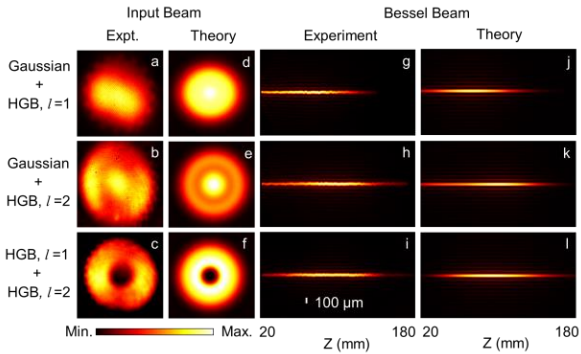

Fig. 4. (a-c) Experimental and (d-f) theoretical spatial intensity distribution of the superposed beams before the axicon. (g-i) Experimental and (j-l) theoretical intensity distribution of the generated Bessel beams along the propagation direction.

To verify the increase in the peak intensity of the Bessel beam due to the decrease in its range with the order of input HGBs, we have studied the frequency doubling characteristics. Keeping the total power of the input beam before the axicon at 5 W, we have measured the SHG power of the Bessel beam from 30 mm long MgO:CLT and 50 mm long MgO:CLN crystals. The results are shown in Fig. 5. As evident from Fig. 5(a), the Bessel beam SHG power (solid dots) of the MgO:CLT increases from 8 mW for the input Gaussian beam to 21 mW for input HGB of order, $l$= 3. Such increase in the SHG power for a fixed pump power and crystal parameters confirms the increase in peak intensity of the Bessel beam with the increase in the order of HGBs. However, in case of 50 mm long MgO:CLN crystal, the SHG power of the Bessel beam decreases from 272 mW for the input Gaussian beam to 227 mW for input HGB of order, $l$= 3. Such decrease in the SHG power can be attributed to the decrease in the effective interaction length of the 50 mm long MgO:CLN crystal due to the lower range of the Bessel beam with the order of HGBs (see Fig. 2). We have also measured the SHG power of the Bessel beam generated from superposed beams. As evident from Fig. 5(b), the Bessel beam SHG power from the 50 mm long MgO:CLN crystal increases for the superposed beams (Gaussian, and HGB, $l$= 1), (HGB, $l$= 1, and HGB, $l$= 2) and (Gaussian, and HGB, $l$= 2). Such increase in the SHG power for a fixed pump power and crystal parameters can be attributed to the increase in the effective nonlinear interaction length of the crystal due to the increase in Bessel beam range. In this experiment, we have used a beam splitter to combine the beams from arm-1 and arm-2 to maintain the beams in same polarization state. The phase matching temperatures are measured to be 44.7° C and 50.9° C for MgO:CLT and MgO:CLN crystals respectively. The far-field intensity distribution of the fundamental and SHG Bessel beams are shown in inset of Fig. 5(a) and Fig. 5(b) respectively.

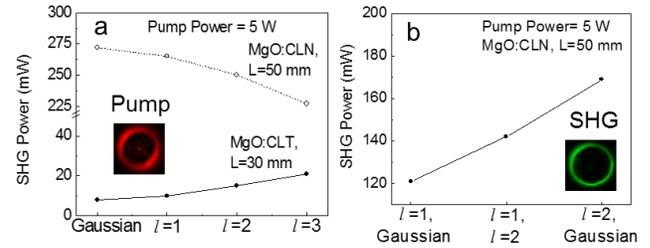

Fig. 5. (a) Variation of SHG power of MgO:CLT and MgO:CLN crystals for input Bessel beam generated from Gaussian beam and HGBs of orders, $l$= 1, 2, and 3. (b) Dependence of Bessel beam SHG power of 50 mm long MgO:CLN crystal on the superposed beam before axicon. (Inset) Far-field intensity distribution of pump and SHG Bessel beams. Lines are guide to eye.

In conclusion, we have proposed and experimentally verified the generation of high power, segmented zero-order Bessel beams of smooth axial intensity distribution and tunable range. Varying the order of the HGBs, $l$=1, 2 and 3, we have generated zero-order Bessel beam with range of 76 mm, 69 mm, and 64 mm and peak position at a distance 62 mm, 74 mm, and 83 mm away from the axicon respectively. We have observed the increase in peak intensity of the zero-order Bessel beam by 44%, 66% and 79% for HGB order, l=1, 2, and 3, respectively, as compared to peak intensity of the Bessel beam generated from Gaussian beam. Using the superposition of HGB of different orders and Gaussian beam we have increased the range up to 151 mm. The increase of peak intensity and the range of the Bessel beam is confirmed with single-pass SHG in nonlinear crystals of different lengths.